\documentclass[aps,prl,twocolumn,floats,amsmath,amssymb,showpacs,floatfix]%
{revtex4}
\usepackage{graphicx}
\usepackage{dcolumn}

\begin{document}

\title{Poincar\' e Indices of Rheoscopic Visualisations}
\author{Vlad Bezuglyy$^{1}$, Bernhard Mehlig$^{2}$ and Michael Wilkinson$^{1}$}
\affiliation{ $^{1}$Department of Mathematics and Statistics, The Open
University, Walton Hall,
Milton Keynes, MK7 6AA, England \\
$^{2}$Department of Physics, G\"oteborg University, 41296
Gothenburg, Sweden \\}

\begin{abstract}
Suspensions of small anisotropic particles, termed \lq rheoscopic fluids', are used for flow visualisation. By illuminating the fluid with light of three different colours, it is possible to determine Poincar\'e indices for vector fields formed by the longest axis of the particles. Because this vector field is non-oriented, half-integer Poincar\' e indices are possible, and are observed experimentally. An exact solution for the direction vector appears to preclude the existence of topological singularities. However, we show that upon averaging over the random initial orientations of particles, singularities with half-integer Poincar\'e index appear. We describe their normal forms.
\end{abstract}



\maketitle

\begin{figure}[t]
\centerline{\includegraphics[width=8.cm]{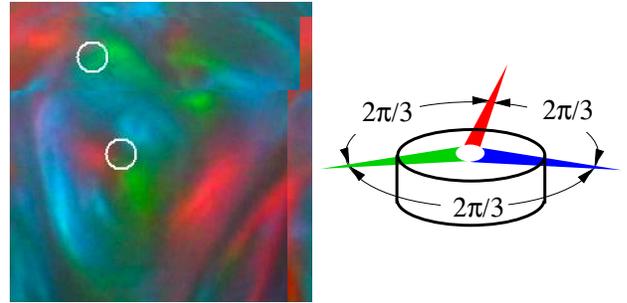}}
\caption{\label{fig: 1} ({\bf a}) Shows textures obtained by illuminating the surface  of randomly stirred rheoscopic fluid with diffuse red, green, and blue light sources, as illustrated in ({\bf b}). The circles in ({\bf a}) indicate the positions of singularities.}
\end{figure}

{\it 1. Introduction}.  Fig.~\ref{fig: 1}{\bf a} is a photograph of the surface of a randomly stirred fluid. The fluid is a suspension of elongated microscopic reflective particles, illuminated by red, green and blue lights (R,G,B) coming from three different directions, as illustrated in
Fig.~\ref{fig: 1}{\bf b}. Two circles indicate points with the property that, on traversing a small clockwise circuit around this point, we encounter the primary colours without repetition (in one case R-G-B, in the other R-B-G). At first sight the existence of such points seems unremarkable, but we argue below that it is indicative of a singularity in the direction field of the particles which appears to be forbidden. Our letter explains why such singularities are nevertheless observed, and classifies their normal forms.

The fluid used to produce Fig.~\ref{fig: 1}{\bf a} was a commercially available {\em rheoscopic} fluid, which is used (diluted in water) for flow visualisation \cite{Mat+84}. It contains microscopic rod-like crystals. It is desirable to understand what aspects of the fluid motion are revealed by the rheoscopic fluid (various aspects are discussed in \cite{Mal+91,Sze+91,Sze93,Shi+97,Gau+98}). In order to pose the questions in the simplest setting, we consider incompressible flow in two dimensions, before finally considering the three-dimensional case. The reflective elements are assumed to be rod-like axisymmetric crystals which are very short compared to the characteristic length scale of the velocity field of the flow, $\mbox{\boldmath$v$}(\mbox{\boldmath$r$},t)$.

In the following we show that Fig.~\ref{fig: 1}{\bf a} demonstrates the existence of singularities in the direction field which have a half-integer Poincar\' e index. (The Poincar\' e index is a topological invariant. For a vector field in the plane, the Poincar\' e index of a closed curve is the number of $2\pi$ clockwise rotations of the vector field as the curve is traversed, also clockwise. Curves with a non-zero Poincar\'e index encircle a singularity of the field.)

\begin{figure}
\centerline{\includegraphics[width=6.cm]{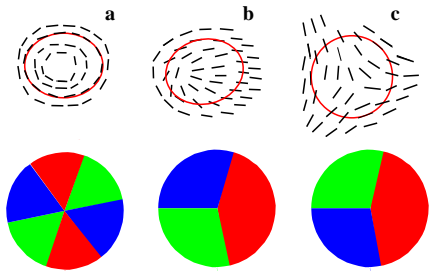}}
\caption{\label{fig: 2} ({\bf a}) Shows the direction field around a vortex
with Poincar\'e index $1$ and the corresponding colour texture
cycling through ${\rm R}\to{\rm G}\to {\rm B} \to {\rm R}\to{\rm G}\to {\rm B}$.
({\bf b}), ({\bf c}) Show the same for a {\em core} singularity with index $\frac{1}{2}$ and for a {\em delta} singularity with index $-\frac{1}{2}$ respectively.}
\end{figure}

There are two reasons why the appearance of such singularities is unexpected. Intuitively, it might be expected that the long axis of the particles will align with streamlines of the flow. Streamlines around a vortex or about a saddle point have Poincar\' e index of $\pm 1$. However, singularities of this type would result in the particles reflecting each colour twice upon making a circuit (see Fig.~\ref{fig: 2}{\bf a}), which is not consistent with Fig.~\ref{fig: 1}{\bf a}. A more persuasive argument is that the exact solution (given below) of the equation of motion for the axis of the crystals shows that the Poincar\'e index of this vector field is invariant, which seems to preclude the emergence of patterns such that shown in Fig.~\ref{fig: 1}{\bf a}. However, to compute the intensity of reflected light one must average over the random initial particle orientations. We show that this gives rise to an {\em order-parameter} field which does exhibit singularities. We remark that, while the use of colours to enhance rheoscopic visualisations was previously demonstrated in \cite{Tho+99}, the technique was not used to reveal singularities.

The axial direction field of the rod-like particles is non-orientable (that is, the sign of the vector is irrelevant). This allows other types of singularity, such as those shown in Figs.~\ref{fig: 2}{\bf b} and {\bf c}, which have a half-integer Poincar\'e index, and which are consistent with the particles reflecting each colour only once upon traversing a closed curve. These singularities have not previously been considered in fluid dynamical problems, although they are seen in fingerprints (where they are known as the {\em core} and {\em  delta}, respectively \cite{Hen00}). Fig.~\ref{fig: 1}{\bf a} is evidence that they are present in our experiment.

2. {\sl Equation of motion and its solution}. The equation of motion for an ellipsoid of revolution in a flow at low Reynolds number was obtained by Jeffrey \cite{Jef22}, and Bretherton \cite{Bre62} showed that the same equation applies to a general axisymmetric body. In the limit of rod-like particles, a unit vector ${\bf n}(t)$ in the direction of the axis of the microscopic particles satisfies
\begin{equation}
\label{eq: 1} \dot{\bf n}={\bf A}{\bf n}-({\bf
n}\cdot {\bf A}{\bf n}){\bf n}
\end{equation}
where ${\bf A}(\mbox{\boldmath$r$}(t),t)$ is the velocity-gradient tensor
at the centre $\mbox{\boldmath$r$}(t)$ of the particle, with elements $A_{ij}=\partial v_i/\partial r_j$. The particle position  advected by the flow, $\dot{\mbox{\boldmath$r$}}=\mbox{\boldmath$v$}(\mbox{\boldmath$r$},t)$.

An exact solution of the equation of motion (\ref{eq: 1}) is obtained from a matrix ${\bf M}(t)$ which is obtained by integrating the linear differential equation
\begin{equation}
\label{eq: 2}
\dot {\bf M}={\bf A}(\mbox{\boldmath$r$}(t),t){\bf M}
\end{equation}
where $\mbox{\boldmath$r$}(t)$ is the trajectory of the centre of the rod. This matrix is the {\em monodromy matrix} describing the evolution of the infinitesimal separation $\delta \mbox{\boldmath$r$}(t)$ of neighbouring points in the flow: we have $\delta \mbox{\boldmath$r$}(t)={\bf M}(t,t_0)\delta \mbox{\boldmath$r$}(t_0)$, where $\delta \mbox{\boldmath$r$}(t_0)$ is the initial infinitesimal separation at time $t_0$. The initial condition for  equation (\ref{eq: 2}) at time $t_0$ is ${\bf M}(t,t_0)={\bf I}$, where ${\bf I}$ is the identity matrix. Now if ${\bf n}_0$ is the initial direction of the rod at time $t_0$, the direction at time $t$ is given by
\begin{equation}
\label{eq: 3}
{\bf n}(t)={\mbox{\boldmath$d$}(t)}\big/{\vert \mbox{\boldmath$d$}(t)\vert}
\ ,\ \ \
\mbox{\boldmath$d$}(t)={\bf M}(t,t_0){\bf n}_0 \ .
\end{equation}
This solution was first given by Szeri \cite{Sze93}. We are interested in the vector field of the rod orientations at position $\mbox{\boldmath$r$}$ and time $t$. The matrix ${\bf M}$ then depends upon position as well as time, and we write ${\bf M}(\mbox{\boldmath$r$},t,t_0)$ for the monodromy matrix of a trajectory which reaches $\mbox{\boldmath$r$}$ at time $t$, starting from $\mbox{\boldmath$r$}_0$ at time $t_0$. It follows from (\ref{eq: 3}) that the vector field of rod orientations is
\begin{equation}
\label{eq: 4}
{\bf n}(\mbox{\boldmath$r$},t)=\frac
{{\bf M}(\mbox{\boldmath$r$},t,t_0){\bf n}_0(\mbox{\boldmath$r$}_0)}
{\vert{\bf M}(\mbox{\boldmath$r$},t,t_0){\bf n}_0(\mbox{\boldmath$r$}_0)\vert}
\ .
\end{equation}

The monodromy matrix ${\bf M}(\mbox{\boldmath$r$},t,t_0)$ is a smooth function of the final position of the trajectory, $\mbox{\boldmath$r$}$. The solution (\ref{eq: 4}) can therefore only be discontinuous if the initial direction field is discontinuous, or if the denominator $\vert {\bf M}{\bf n}_0\vert$ is equal to zero, which is not possible because ${\rm det}({\bf M})=1$. If the initial direction vector field ${\bf n}_0(\mbox{\boldmath$r$})$ is non-singular, we therefore conclude that the direction field ${\bf n}(\mbox{\boldmath$r$},t)$ remains non-singular for all times. Because the vector field generated by (\ref{eq: 4}) is smooth, the Poincar\'e index of this field is zero for any closed curve.

In an earlier paper \cite{Wil+08} we discussed the textures formed by rods-like particles in a complex flow for a specified initial direction field ${\bf n}_0(\mbox{\boldmath$r$})$. Here we deal with the more complex case where we must average over the random initial orientation. We show below that this leads to singularities of the orientation field.

3. {\sl An order parameter for rheoscopic fluids}. Initially, at time $t_0$, the rod-like particles in a rheoscopic fluid are randomly oriented, due to the effects of Brownian motion. We must therefore consider the distribution of rod directions generated by the solution (\ref{eq: 4}) at each point in the flow. This can be described by a probability density $P(\theta)$ for the rod orientation angle $\theta$ (satisfying $P(\theta+\pi)=P(\theta)$, because the rods are non-oriented). This probability density depends upon both position and time. Specifying the angle distribution at each point in space and time would provide too much information to be a useful description. It is therefore desirable to map this distribution $P(\theta,\mbox{\boldmath$r$},t)$ to an order parameter vector $\mbox{\boldmath$\zeta$}(\mbox{\boldmath$r$},t)$. The direction of this non-oriented vector should represent the predominant direction of the rods, and its magnitude should indicate the degree of ordering (with $\vert\mbox{\boldmath$\zeta$}\vert =1$ when the rods are all in the same direction, and $\vert\mbox{\boldmath$\zeta$}\vert=0$ when their angular distribution is isotropic).

The initial direction vector ${\bf n}_0$ in (\ref{eq: 2}) is a random vector, uniformly distributed about the unit circle. The vector $\mbox{\boldmath$d$}(t)={\bf M}(t,t_0){\bf n}_0$ is therefore distributed about an ellipse. If the unit circle is represented as $\mbox{\boldmath$x$}\cdot\mbox{\boldmath$x$}=1$, the ellipse upon which $\mbox{\boldmath$d$}$ lies is represented by the equation
\begin{equation}
\label{eq: 5}
\mbox{\boldmath$x$}\cdot {\bf K}\mbox{\boldmath$x$}=1
\ ,\ \ \
{\bf K}=({\bf M}{\bf M}^{\rm T})^{-1}
\ .
\end{equation}
This ellipse has its longest axis aligned along a direction $\bar \theta$, which is in the direction of the eigenvector corresponding to the largest eigenvalue of ${\bf M}\,{\bf M}^{\rm T}$. Its aspect ratio $\nu$ is the square root of the ratio of the eigenvalues of ${\bf M}{\bf M}^{\rm T}$ (we choose to consider $\nu \ge 1$). It is natural to define the magnitude of the order parameter to be a function of $\nu $ which interpolates between zero (when $\nu=1$) and unity (as $\nu\to\infty$). In a later paper we shall argue that the following expression is the most natural definition for the order parameter:
\begin{equation}
\label{eq: 6}
\mbox{\boldmath$\zeta$}=\frac{\nu-1}{ \nu +1}\,{\bf n}(\bar\theta)\,,
\end{equation}
where ${\bf n}(\theta)$ is a non-oriented unit vector with angle $\theta$.

4. {\sl Relation between order parameter and light scattering}. The colour which is reflected by the rheoscopic fluid in the experiment illustrated in Fig.~\ref{fig: 1} may be related to the order parameter. The details of this relation depend on the ratio of the length of the rod-like particles in the rheoscopic fluid to the wavelength of light, and upon their surface roughness. For illustration we discuss the simplest case, where the rods are short compared to the wavelength of light. In this limit the scattered intensity from a rod is proportional to the square of the projected cross-section, so that a rod at angle $\theta$ scatters light from a source which is perpendicular to the direction $\phi$ with an intensity proportional to $\cos^2(\theta-\phi)$. The combined effect of scattering from the three light sources results in additive colour mixing, so that at any given point the reflected colour $C$ is a weighted combination of red, green and blue ($R$, $G$, $B$) of the form
\begin{equation}
\label{eq: 7}
C=I(0)\,R+I(2\pi/3)\,G+I(4\pi/3)\,B
\end{equation}
where $I(\phi)$ is the average of $\cos^2(\theta-\phi)$ over the rod orientation:
\begin{equation}
\label{eq: 8}
I(\phi)=\int_0^{2\pi}\!\!\!\!{\rm d}\theta\ P(\theta) \cos^2(\phi-\theta)
\ .
\end{equation}
After a lengthy but elementary calculation we obtained the following expression for the reflected colour in terms of the parameters ($\bar\theta$, $\nu$) of the order parameter:
\begin{eqnarray}
\label{eq: 9}
C&\!=\!&\frac{1}{4(\nu+1)}\bigl\{
4R[\nu\cos^2 \bar\theta+\sin^2 \bar\theta]\\
&+&G[2 (1\!-\!\nu) \cos ^2 \bar\theta
-2\sqrt{3}(\nu\! - \!1) \sin \bar\theta \cos \bar\theta + 3\nu\! +\!1]
\nonumber\\
&+&B[2 (1\!-\!\nu) \cos ^2 \bar\theta+2\sqrt{3} (\nu\!  - \!1)
\sin \bar\theta \cos \bar\theta + 3\nu \!+\!1]
\bigr\}\ .\nonumber
\end{eqnarray}

\begin{figure}
\centerline{\includegraphics[width=9.cm]{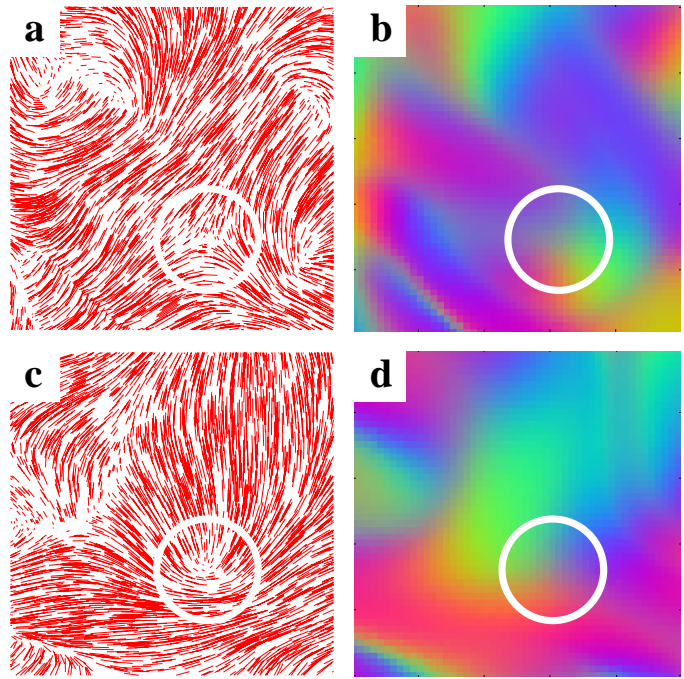}}
\caption{\label{fig: 3} ({\bf a}) Shows the order parameter field for a random flow. ({\bf b}) Shows the same image colour coded using equation (\ref{eq: 9}) to indicate the additive colour mixing of light from three different sources. The circle marks the position of a {\em delta} singularity;
({\bf c}) and ({\bf d}) the same, but for a {\em core} singularity.}
\end{figure}

Figs.~\ref{fig: 3}{\bf a} and {\bf c} show the order parameter obtained from (\ref{eq: 4}) and (\ref{eq: 6}) for small rods moving in a random flow field (we used the same model as in \cite{Wil+08}), starting from a uniform distribution of angles. In the vicinity of zeros of the order parameter, there are patterns which resemble the {\em delta} (Fig.~\ref{fig: 3}{\bf a})
and {\em core} (Fig.~\ref{fig: 3}{\bf c}) singularities illustrated in
Figs.~\ref{fig: 2}{\bf c} and {\bf b}.

The remaining panels of Fig.~\ref{fig: 3} ({\bf b} and {\bf d})
show the light reflected from three coloured sources, computed using equation (\ref{eq: 9}). (The additive colour mixing for this figure was performed using MatLab.). These singularities have a Poincar\' e index of $\pm \frac{1}{2}$, and they are therefore consistent with the experimental result shown in Fig.~\ref{fig: 1}{\bf a}. In the remainder of this letter we discuss how the occurrence of these structures can be understood, and the possibility that additional types of singularity might appear if the flow is three-dimensional.

5. {\sl Singularities of the order parameter}. If zeros of the vector order parameter field exist, the Poincar\'e index of a curve may be non-zero even though the order parameter depends smoothly upon position. A singularity where the order parameter is equal to zero occurs where the ellipse upon which the vector $\mbox{\boldmath$d$}(t)={\bf M}(t,t_0){\bf n}_0$ lies degenerates to a circle. Thus zeros of the order parameter occur when ${\bf M}$ is a rotation matrix.

First we consider whether such singularities are generic. Because the two-dimensional flow is area-preserving, the $2\times 2$ matrix ${\bf M}$ satisfies ${\rm det}({\bf M})=1$. This matrix can be written in a form determined by three parameters $\lambda$, $\kappa$, $\chi$:
\begin{equation}
\label{eq: 10}
{\bf M}={\bf D}(\lambda,\lambda^{-1})\,{\bf S}(\kappa)\,{\bf O}(\chi)
\end{equation}
where ${\bf D}$, ${\bf S}$ and ${\bf O}$ are respectively diagonal, shear and rotation matrices:
\begin{eqnarray}
\nonumber
{\bf D}(\lambda_1,\lambda_2)&=&
\left(\begin{array}{cc}
\lambda_1 & 0         \cr
0         & \lambda_2
\end{array}\right)
\ \,,\ \
{\bf S}(\kappa)=\left(\begin{array}{cc}
1 & \kappa  \cr
0 & 1
\end{array}\right)\,,\\
\label{eq: 11}
&&\mbox{}\hspace*{-5mm}{\bf O}(\chi)=\left(\begin{array}{cc}
\cos\chi  &\sin\chi \cr
-\sin\chi &\cos\chi
\end{array}\right)\ .
\end{eqnarray}
The singularity occurs when $\lambda\!=\!1$ and $\kappa\!=\!0$ in (\ref{eq: 10}) (with no condition upon $\chi$), which is realised upon varying two parameters. Since the coordinate space is two-dimensional, the singularities occur at isolated points in the plane.

\begin{figure}
\centerline{\includegraphics[width=9.cm]{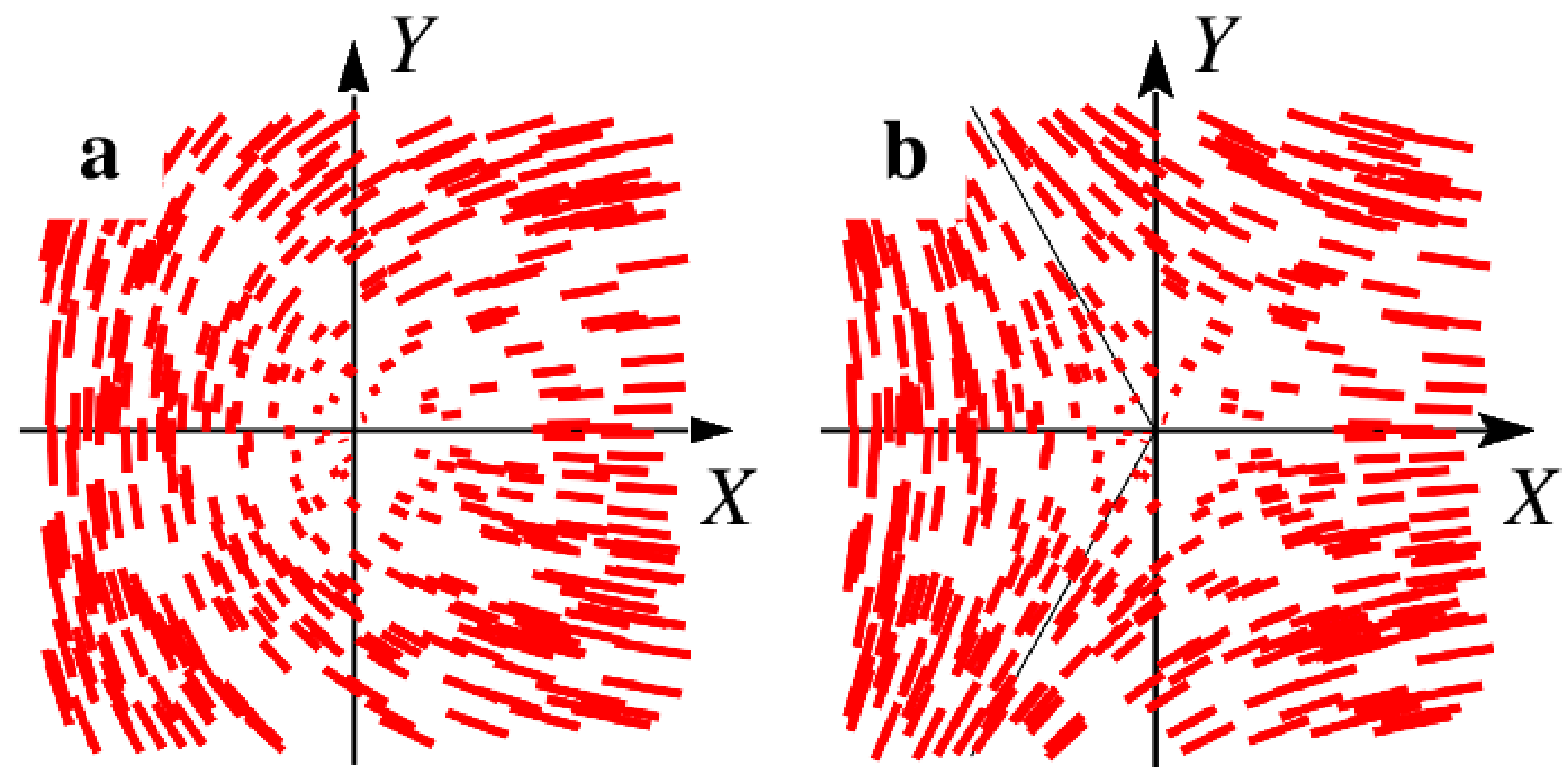}}
\caption{\label{fig: 4} ({\bf a}) Shows the order parameter field $\mbox{\boldmath$\zeta$}(\mbox{\boldmath$R$})$ for the canonical singularity of eq.~(\ref{eq: 12}), with a positive sign multiplying $Y$. This is a {\em core} singularity. ({\bf b}) Shows the order parameter field when the negative sign is chosen. This is a {\em delta} singularity. }
\end{figure}

Having identified the condition defining the singular point, we now turn to consider the form of the order parameter field in its vicinity. To leading order, generically the parameters $\lambda$ and $\kappa$ depend linearly on position in the vicinity of a singular point at which the monodromy matrix becomes a pure rotation. Let us assume that there is a singular point at  $\mbox{\boldmath$r$}_0=(x_0,y_0)$. In the vicinity of this point there exists a coordinate system $\mbox{\boldmath$R$}=(X,Y)$ such that the monodromy matrix is in the normal form
\begin{equation}
\label{eq: 12}
{\bf M}(\mbox{\boldmath$R$})={\bf D}(1+\tfrac{1}{2}X,1-\tfrac{1}{2}X)\,{\bf S}(\pm Y)\,{\bf O}(\chi)+{\cal O}(\mbox{\boldmath$R$}^2)
\ .
\end{equation}
The local coordinate system is related to $\mbox{\boldmath$r$}$ by a linear transformation: $\mbox{\boldmath$r$}-\mbox{\boldmath$r$}_0={\bf T}\mbox{\boldmath$R$}$, where ${\bf T}$ is a $2\times 2$ matrix and the sign in (\ref{eq: 12}) is chosen so that ${\bf T}$ is a non-inverting transformation (${\rm det}({\bf T})>0$). The vector order parameter for the normal form is plotted in Fig.~\ref{fig: 4} for both choices of the sign in (\ref{eq: 12}), and it can be seen that these two patterns resemble the core and delta singularities of fingerprints (as described in \cite{Hen00}). It is instructive to analyse the behaviour of the order parameter in the vicinity of these singularities. Retaining only the leading order terms in $X$ and $Y$ from (\ref{eq: 12}), we find
\begin{equation}
\label{eq: 13}
{\bf M}\,{\bf M}^{\rm T}=\left(\begin{array}{cc}
1+X & \pm Y \cr
\pm Y & 1-X \cr
\end{array}\right)+{\cal O}(R^2)
\end{equation}
where $R=\sqrt{X^2+Y^2}$. The eigenvalues of ${\bf M}{\bf M}^{\rm T}$ are $\lambda_{\pm}=1\pm R+{\cal O}(R^2)$, so that $\nu = \sqrt{\lambda_+/\lambda_-}=R+{\cal O}(R^2)$. Writing $(X,Y)=(R\cos\Theta,R\sin\Theta)$ and $(\cos\Phi,\sin\Phi)^{\rm T}$ for the eigenvector of ${\bf M}{\bf M}^{\rm T}$ corresponding to $\lambda_+$, we find $\Phi=\pm\frac{1}{2}\Theta$, so that
\begin{equation}
\label{eq: 14}\mbox{\boldmath$\zeta$}(R,\Theta)=R\,{\bf n}(\pm \tfrac{1}{2}\Theta)+{\cal O}(R^2)\ .
 \end{equation}
If the positive sign is chosen, $\mbox{\boldmath$\zeta$}$ points in the radial direction ($\Phi=\Theta\,{\rm mod}\pi$) for only one ray ($\Theta=0$). For the negative sign, $\mbox{\boldmath$\zeta$}$ points radially outwards along three rays ($\Theta=0$, $\pm 2\pi/3$). These properties are characteristic of the core and delta singularities, respectively.

6. {\sl Three-dimensional flows}. Thus far, we have considered two-dimensional flows. In the experiment illustrated in Fig.~\ref{fig: 1}{\bf a}, the depth of the rheoscopic fluid was a few millimetres, and comparable to the scale size of structures in the texture shown in Fig.~\ref{fig: 1}{\bf a}, which shows a region which is a few centimetres across.  We should therefore consider what additional structures might arise when the flow is three-dimensional. In a three dimensional flow the direction vector ${\bf n}(\mbox{\boldmath$r$},t)$ covers a sphere rather than a circle (it remains non-oriented). The optical depth was very small, so that light was reflected from a thin layer of fluid just below the surface. It is only the direction of the projection of ${\bf n}(\mbox{\boldmath$r$},t)$ in the plane of the fluid surface which determines the colour of the scattered light. This projected vector field, ${\bf n}_{\rm p}(\mbox{\boldmath$r$},t)$, can have simple zeros if there are positions where the rods point out of the surface of the liquid. The projected vector field ${\bf n}_{\rm p}(\mbox{\boldmath$r$},t)$ then has a simple zero, with Poincar\'e index $+1$. We conclude that if the flow is three dimensional, double rotations of the primary colours (illustrated in Fig.~\ref{fig: 2}{\bf a}) may be observed, as well as single rotations
(as in Figs.~\ref{fig: 2}{\bf b} and {\bf c}).

7. {\sl Concluding remarks}. We have shown that our simple experiment on a rheoscopic fluid reveals singularities of the direction field with Poincar\'e index equal to $\pm \frac{1}{2}$. We defined an order parameter for this field, and showed that its zeros have normal forms which are analogous to the core and delta singularities of fingerprints. The distinctive behaviour of the order parameter in steady flows will be addressed in a longer paper, in preparation.

{\it Acknowledgements}. VB was supported by an Open University postgraduate studentship. BM was supported by the Vetenskapsr\aa{}det and the research initiative \lq Nanoparticles in an interacting environment' at Gothenburg University.

\end{document}